# FIND: an SDR-based Tool for Fine Indoor Localization

Evgeny Khorov, Aleksey Kureev, Vladislav Molodtsov
Institute for Information Transmission Problems, Russian Academy of Sciences, Moscow, Russia
Email: {khorov, kureev, molodtsov}@wireless.iitp.ru

*Abstract*—An indoor localization approach uses Wi-Fi Access Points (APs) to estimate the Direction of Arrival (DoA) of the Wi-Fi signals. This paper demonstrates FIND, a tool for Fine INDoor localization based on a software-defined radio, which receives Wi-Fi frames in the 80 MHz band with four antennas. To the best of our knowledge, it is the first-ever prototype that extracts from such frames data in both frequency and time domains to calculate the DoA of Wi-Fi signals in real-time. Apart from other prototypes, we retrieve from frames comprehensive information that could be used to DoA estimation: all preamble fields in the time domain, Channels State Information, and signal-to-noise ratio. Using our device, we collect a dataset for comparing different algorithms estimating the angle of arrival in the same scenario. Furthermore, we propose a novel calibration method, eliminating the constant phase shift between receiving paths caused by hardware imperfections. All calibration data, as well as a gathered dataset with various DoA in an anechoic chamber and in a classroom, are provided to facilitate further research in the area of indoor localization, intelligence surfaces, and multi-user transmissions in dense deployments [1].

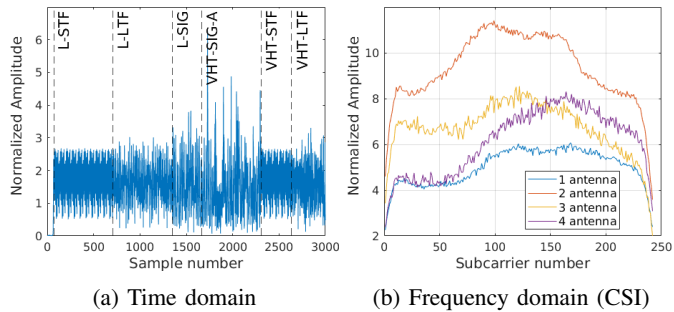

(a) Time domain  (b) Frequency domain (CSI)

Figure 1: Example of obtained information.

## I. INTRODUCTION

The indoor localization problem has attracted much attention recently. For instance, various scenarios require positioning inside buildings, where conventional solutions, such as GPS, are not applicable. Recently, localization has become crucial for a novel paradigm of intelligent surfaces [1]. A possible approach to the indoor localization problem is to use Wi-Fi Access Points (APs), which are currently widespread. Many of them utilize antenna arrays to increase throughput with the MIMO transmissions. Information, e.g., the gain and phase difference, received by these arrays could be used to estimate the Direction of Arrival (DoA) of the signals from client devices. Afterward, such directions from several APs can be combined together to reveal the device location.

Several promising techniques, introduced in papers [2, 3], including machine learning methods in paper [4], show excellent results with decimeter accuracy. However, the problem of the best DoA estimation is still open because of two issues. First, the existing approaches have never been compared with each other *in the same scenarios*. Moreover, they may rely on the different channel information. Some of them [3, 4], operate with Channel State Information (CSI), while others [2] require IQ samples in the time domain. Unfortunately, there are no available datasets containing enough information to compare all the algorithms. Second, the existing approaches are implemented on various hardware platforms with different capabilities, often too poor. Many works [3, 4] exploit Intel CSI Tool, providing CSI for 30 subcarriers out of over a hundred of a 40 MHz frame, while the CSI values for the rest are skipped. A rare exception is a software-defined radio testbed [2] that employs only L-STF, the first field of the Legacy preamble, which is used for start-of-frame detection, for coarse frequency offset compensation, and for setting the receiver's amplifier. Other fields remain unused, while the amount of information extracted from each frame directly affects the quality of DoA estimation.

To solve these issues, we present FIND (the tool for Fine INDoor localization), the 4-channel receiver based on NI USRP-2955, capable of capturing real Wi-Fi frames in the 80 MHz band. It retrieves CSI from all 242 subcarriers, as well as all time-domain IQ samples of the Legacy preamble, needed for backward compatibility, and VHT preamble, its extension for higher data rates. Fig. 1 shows an example of such information. Eventually, we introduce a dataset with all such information at different DoA, which can be used for comparing different DoA estimation algorithms under equal conditions. In addition, we provide supplementary data for calibration by our invented method detailed below. The developed prototype can be used not only for DoA estimation in typical localization systems and emerging reconfigurable intelligent surfaces. In addition, it can also be used for the investigation of the properties of the MIMO channel in Wi-Fi networks.

## II. FIND DESCRIPTION

In our testbed, we use NI USRP-2955 as a receiver connected to the laptop having LabVIEW 2020 software. This equipment allows data receiving and processing in an 80 MHz band from four channels simultaneously. It has an FPGA inside, facilitating the implementation of some parts of Wi-

[1]The research has been done at IITP RAS and supported by the Russian Science Foundation (agreement No 20-19-00788). The obtained dataset is available at http://wireless.iitp.ru/dataset-for-wi-fi-localization/.

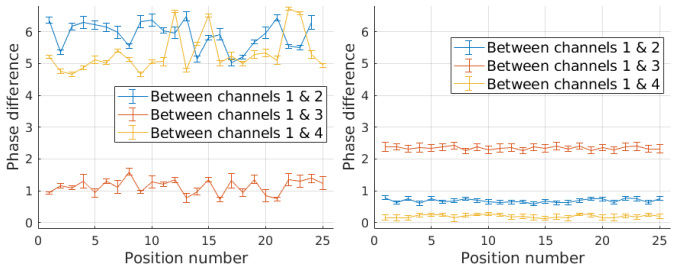

(a) In a classroom  (b) In an anechoic chamber

Figure 2: Phase calibration.

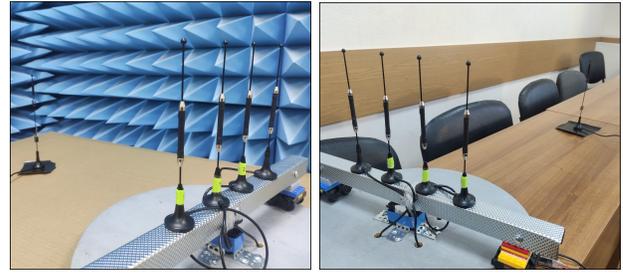

(a) In an anechoic chamber  (b) In a classroom

Figure 3: Photos of the testbed.

Fi PHY needed for frame detection, carrier frequency offset compensation, and CSI extraction. After frame reception, all necessary information is streamed to the host. The receiver's antenna array is located on a rotating platform, which exposed the array at a given angle relative to the transmitter, which is is NI USRP-2945, but any Wi-Fi device can be used.

In practice, even though different channels share the same local oscillator, there is a phase shift between signals arrived from them, which is detrimental to localization accuracy. By channel here, we mean a combination of antenna, feedline, and RF chain. Nonetheless, this phase difference is constant in time and is caused solely by imperfections of components mentioned above, as stated in paper [5]. Therefore, only one-time calibration must be performed to eliminate its impact.

To remove such shifts, the authors of paper [2] proposed to send a reference signal to all channels via splitter and measure the phase difference between them. However, this procedure does not take into account imperfections of feedlines and antennas used in our case.

Another fruitful technique originating from paper [3] is to deploy the receiving antennas at the same distance from the transmitting one and measure the phase in this case. According to the authors of the method, it is necessary and sufficient to install the transmitting antenna closer to the receiving ones in order to mitigate the impact of multipath in the indoor environment. But it proved not to be the case in our scenario, in which the effect of multipath is still significant. We found out that such phase shifts measured in different places in space are non-identical even when antennas almost touch each other. It can be seen from Fig. 2a, where we plot the phase difference between channels in various positions in a classroom.

Thereby, we proposed a new procedure to overcome these problems. We decided to perform calibration in the same way, as in paper [3], but in an anechoic chamber, where the impact of multipath is reduced vastly. Fig. 2b confirms the advantages of our method since the phase shift is almost independent of the position of the antenna array in this case.

### III. GATHERED DATASET

Currently, FIND works with 802.11ac Legacy and VHT preambles, retrieving from them time-domain IQ samples of all fields, CSI from all 242 subcarriers of VHT-LTF field, and signal-to-noise ratio. This data can be used for comparing DoA estimation algorithms requiring different information under equal conditions. Unfortunately, due to hardware issues, it operates at 2.4GHz instead of 5GHz declared in 802.11ac standard. According to the specification, USRP-2955 has a frequency range of up to 6GHz. In spite of this, we found out that the amplifiers of the channels RX 0 RF 0 and RX 1 RF 0 do not work properly at 5GHz. Moreover, this problem is observed in several different USRP-2955. However, in the nearest future, we plan to solve these issues and expand the capabilities of our tool for running at other frequencies and with other preambles, such as HT, HE, and EHT [6]. We are also going to carry out experiments in other scenarios, e.g., in malls. As collected, the additional data will be included in the new versions of the published dataset.

Fig. 1 illustrates the gathered information. It contains the absolute value of time-domain IQ samples of various fields of preambles mentioned above from one antenna (see Fig. 1a), as well as the absolute value of CSI from the VHT-LTF field on all four antennas, which is represented in Fig. 1b. These absolute values are normalized to simplify the presentation.

With the devised testbed shown in Fig. 3, we collect data from over 300.000 frames in a classroom and over 80.000 frames in an anechoic chamber. The receiver and transmitter are placed in different positions in space, while the real DoA is also tracked. We carry out experiments in semi-automatic mode: we manually set the testbed to a certain position, then the platform is rotated at a defined angle, where FIND receives several frames, after which antennas are rotated to the next angle. All gathered data is made open for further development and testing of DoA algorithms.